\documentclass{PoS}
\usepackage{multirow}

\newcommand{\U}[1]{\ensuremath{\mathrm{\ #1}}}

\def\arcmin{\hbox{$^\prime$}}

\usepackage{multicol}

\makeatletter
\renewenvironment{thebibliography}[1]
     {\begin{multicols}{2}[\section*{\refname}]%
      \@mkboth{\MakeUppercase\refname}{\MakeUppercase\refname}%
      \list{\@biblabel{\@arabic\c@enumiv}}%
           {\settowidth\labelwidth{\@biblabel{#1}}%
            \leftmargin\labelwidth
            \advance\leftmargin\labelsep
            \@openbib@code
            \usecounter{enumiv}%
            \let\p@enumiv\@empty
            \renewcommand\theenumiv{\@arabic\c@enumiv}}%
      \sloppy
      \clubpenalty4000
      \@clubpenalty \clubpenalty
      \widowpenalty4000%
      \sfcode`\.\@m}
     {\def\@noitemerr
       {\@latex@warning{Empty `thebibliography' environment}}%
      \endlist\end{multicols}}
\makeatother

\title{VERITAS and Fermi-LAT observations of TeV gamma-ray sources from the second HAWC catalog}

\ShortTitle{ VERITAS and Fermi-LAT observations for new 2HWC sources}
\author{\speaker{Nahee Park} for the VERITAS Collaboration\thanks{veritas.sao.arizona.edu}, \textit{Fermi}-LAT Collaboration, and HAWC collaboration\\
        University of Chicago\\
        E-mail: \email{nahee@uchicago.edu}}
        
\abstract{The HAWC  observatory recently published their second source catalog with 39 very high energy gamma-ray sources based on 507 days of exposure time. Among these, there were 16 sources that are more than one degree away from any known TeV source. We studied 13 of these 16 sources with VERITAS and \textit{Fermi}-LAT data. VERITAS, an array of four imaging atmospheric Cherenkov telescopes observing gamma rays with energies higher than 85 GeV, can provide a more detailed image of the source with much shorter exposure time and with better angular resolution. With \textit{Fermi}-LAT data, we searched for the counterparts at lower energies (E>10 GeV). The exposure of VERITAS varies among the 13 sources, and we used eight years of \textit{Fermi}-LAT data. VERITAS found weak gamma-ray emission in the region of PWN~DA~495 coinciding with 2HWC J1953+294 in this follow-up study. We will discuss the results especially focusing on the PWN~DA~495 region and the SNR~G~54.1+0.3 region where \textit{Fermi}-LAT detected a GeV counterpart of SNR~G~54.1+0.3, a known TeV source detected by both VERITAS and HAWC.}

\FullConference{35th International Cosmic Ray Conference --- ICRC2017\\
		10--20 July, 2017\\
		Bexco, Busan, Korea}

\begin{document}
\section{Introduction}
The current generation of gamma-ray observatories covers nearly seven orders of magnitude in energy. To cover this wide energy range, the observatories are utilizing different techniques. Space based satellite experiments, such as \textit{Fermi}-LAT, can measure the gamma rays from several tens of MeV by directly collecting the information from gamma-ray interactions within the detector. Satellite experiments have very wide fields of view, can cover the entire sky with very small dead time, and are almost background-free. 
However, due to the limited effective area, which is on the order of 1 $\mathrm{m}^{2}$, the peak sensitivity is around a few GeV. Ground-based detector arrays are best suited to observe gamma rays with energies higher than several hundreds of GeV. They can provide a much larger effective area, on the order of $10^{5}$ $\mathrm{m}^{2}$. Ground-based imaging atmospheric Cherenkov telescope (IACT) arrays, such as VERITAS~\cite{2002APh....17..221W}, provide the best instantaneous sensitivity from energies higher than several hundreds of GeV due to excellent rejection of the cosmic-ray background. However, the observatories have limited duty cycles ($<$20\%), because they require clear dark sky observing conditions, and their fields of view are relatively small ($\lesssim$5$^{\circ}$ diameter). Air shower arrays for gamma-ray observation, such as the High Altitude Water Cherenkov (HAWC) observatory, on the other hand, can provide continuous observation with a field of view covering $\sim$15\% of the sky. The energy and angular resolution provided by these three techniques also has many complementary aspects. 
For example, IACTs can provide better angular resolution in the energy range of several hundreds of GeV up to tens of TeV, compared to air shower arrays. This allows IACTs to study the detailed morphology of an angularly extended source. But, the study of gamma-ray emission from a region that is comparable to or larger than their field of view is generally difficult for IACTs. Thus, the best way to understand the astrophysical gamma-ray sources in detail is by combining observations from multiple observatories.  

HAWC, which recently completed full detector deployment in 2015, is the most sensitive air shower array for gamma-ray observations. HAWC recently published their second source catalog (2HWC) based on 507 days of exposure time from November 2014 to May 2016~\cite{2017arXiv170202992A}. 2HWC contains 39 sources. Of these, 16 sources are at least 1$^\circ$ away from any known TeV gamma-ray emitting source, forming a newly detected TeV source population. In this paper, we describe our effort to understand the properties of some of these new TeV gamma-ray sources by using VERITAS and \textit{Fermi}-LAT data.
 
VERITAS is an array of four IACTs located at the Fred Lawrence Whipple Observatory in southern Arizona~\cite{2002APh....17..221W}. With a field of view of 3.5$^{\circ}$, VERITAS is designed to detect gamma rays from an energy of 85 GeV to energies higher than 30 TeV. VERITAS can detect a point source with 1$\%$ Crab Nebula strength within 25 hours, and has an angular resolution better than 0.1$^{\circ}$ at 1 TeV. 

The LAT is a high-energy gamma-ray telescope which detects photons in the energy range between 20 MeV to higher than 500 GeV~\cite{2009ApJ...697.1071A}. The latest version of event reconstruction, dubbed Pass 8 \cite{2013arXiv1303.3514A}, offers a greater acceptance and an improved PSF compared to previous LAT analyses. With Pass 8 data, the 68\% containment radius of the LAT is less than 0.2$^\circ$ above 10 GeV. 

\section{VERITAS Observations}
We searched for the 16 newly detected HAWC sources in VERITAS archival data collected from 2007 to 2015. We selected data if the pointing of the telescopes was offset by less than 1.5$^{\circ}$ from the locations of these HAWC sources. We found that VERITAS observed locations coincident with 11 out of 16 HAWC sources. In addition to the archival data, VERITAS observed a subset of the HAWC sources during the 2015--2016 and 2016--2017 seasons.  
Combining the archival and new data sets, VERITAS observed a total of 13 new HAWC sources out of 16. After data quality selections, a total of 169 hours of data were analyzed for the study. The exposure time for each source varies from 1.3 hours to 46 hours.

\section{Analyses}
The VERITAS analysis begins with standard calibration and image cleaning procedures, followed by image parameterization. Selection cuts are applied to discriminate gamma-ray-initiated events from background cosmic-ray-initiated air showers. A detailed description of the VERITAS data analysis procedure can be found in~\cite{2008ICRC....3.1325D}. The choice of the optimum gamma-ray selection cuts depends on the assumed strength, spectral energy distribution (SED), and spatial extension of the source candidate. Because the peak sensitivity of HAWC is located in the multi-TeV energy range, the most sensitive gamma-ray selection cuts are optimized for weak sources ($\sim$1\% of the steady Crab Nebula flux) with hard spectral indices ~\cite{2015arXiv150807070P}. We apply these cuts if the VERITAS exposure time for the source is larger than $10\U{hours}$. For sources with less than $10\U{hours}$ of exposure time, we choose a less strict set of cuts suitable for stronger sources ($3\%\U{Crab}$) because a weaker source would be below the sensitivity of the more stringent cuts for such a short exposure. We apply two sets of angular cuts for this study: one for a point-like source search and the other for a moderately extended source search. The moderately extended source search used the angular cut value of 0.23$^{\circ}$. The results described here have been confirmed by two independent analysis chains. 


For the \textit{Fermi}-LAT analysis, we analyzed 8.5 years of LAT data from 2008 August to 2017 February. We used the latest version of the event-level analysis, also known as Pass 8\cite{2013arXiv1303.3514A}. 
To search for the counterparts of HAWC sources in the GeV energy range, we selected events with energies higher than $10\U{GeV}$ from LAT data. By limiting the analysis to such high energies, we can avoid contamination from gamma-ray pulsars in the Galactic plane as well. The region within 10$^{\circ}$ of the HAWC source is fit using a binned likelihood method. In addition to the point source search, we also fit a uniform disk at the location of HAWC source. We fit the localization and extension of the source while searching for a possible LAT counterpart.  
For both VERITAS and \textit{Fermi}-LAT analyses, we calculate 99$\%$ confidence upper limits at the location of HAWC source. 

\section{Results}
\begin{figure}[ht!]
  \centering
  \includegraphics[scale=0.8]{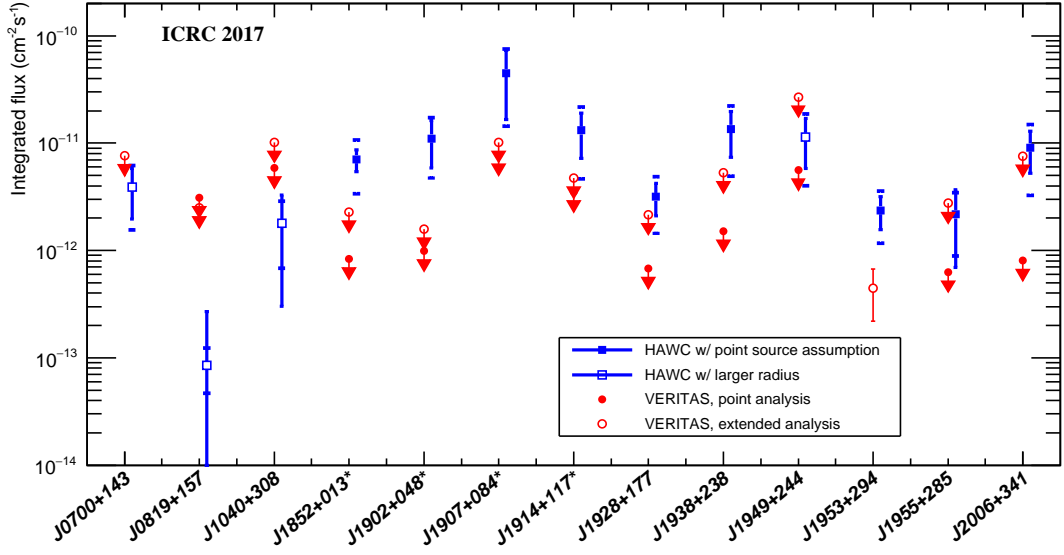}
\caption{Flux comparison between HAWC and VERITAS in the VERITAS energy range. The error bars for the HAWC flux are statistical uncertainties derived from the propagation of statistical uncertainties on the normalization factors and indices. The error bracket of the HAWC flux is the systematic uncertainties. Four sources among 13 selected HAWC sourcesare detected by HAWC in their extended source search.} 
\label{Fig:FluxComparisons}
\end{figure}


\textit{Fermi}-LAT did not detect counterparts for any of the 13 HAWC sources considered in this study, in either the point source or extended source searches. However, we detected the LAT counterpart of a known TeV source, SNR G54.1+0.3, which was located near the newly detected 2HWC source 2HWC J1928+177. No gamma-ray source was detected by VERITAS for 12 of the selected 2HWC sources, but weak emission was detected in the region of 2HWC J1953+294.

To compare the VERITAS upper limits with the flux measurement of HAWC, we calculate the integrated flux from each source in the VERITAS energy range, using the spectral information measured by HAWC. The result is shown in Figure~\ref{Fig:FluxComparisons}. Error bars for the HAWC flux estimates were derived with error propagation using the statistical errors on the flux normalization factors at 7 TeV and the spectral indices. The systematic errors on the HAWC flux, shown as brackets, were calculated with a flux normalization error of 30\% and an index error of 0.2~\cite{2017arXiv170101778A}. 

For the point source search, the 99\% upper limits of VERITAS are lower than the expected fluxes, estimated by using HAWC's spectral information, for most of the sources except three---2HWC J0819+157, 2HWC J1040+308, and 2HWC J1949+244. The exposure of VERITAS on these three sources is relatively small (1.8$\sim$7 hours), and the upper limits are not strongly constraining. Assuming the statistical uncertainties on HAWC's flux estimation follow a Gaussian distribution and neglecting systematic errors, at the 95$\%$ confidence level we exclude five sources---2HWC~J1852+013*, 2HWC~J1902+048*, 2HWC~J1928+177, 2HWC~J1938+238, and 2HWC~J2006+341---from being point sources with the same power-law spectral energy distribution as measured by HAWC. To explain the disagreement between VERITAS and HAWC measurements, the source must have either a harder spectral index or an extended gamma-ray emitting region in the VERITAS energy range.

For the extended source region analysis, the upper limits of VERITAS are less constraining. However, the VERITAS upper limits with an angular cut value of 0.23$^{\circ}$ are lower than the flux estimated by HAWC for three sources: 2HWC J1852+013$^{\ast}$, 2HWC J1902+048$^{\ast}$, and 2HWC J1907+084$^{\ast}$. The discrepancies between the VERITAS and HAWC measurements are especially large for 2HWC J1852+013$^{\ast}$ and 2HWC J1902+048$^{\ast}$. The measurements for these two sources disagree at a confidence level of greater than 95\%. To satisfy both the VERITAS upper limits and the measured HAWC fluxes, the source extensions must be larger than a radius of 0.23$^{\circ}$ for these sources.

\subsection{SNR G 54.1+0.3 region}
\begin{figure}[h!]
  \centering
  \includegraphics[width=0.5\textwidth]{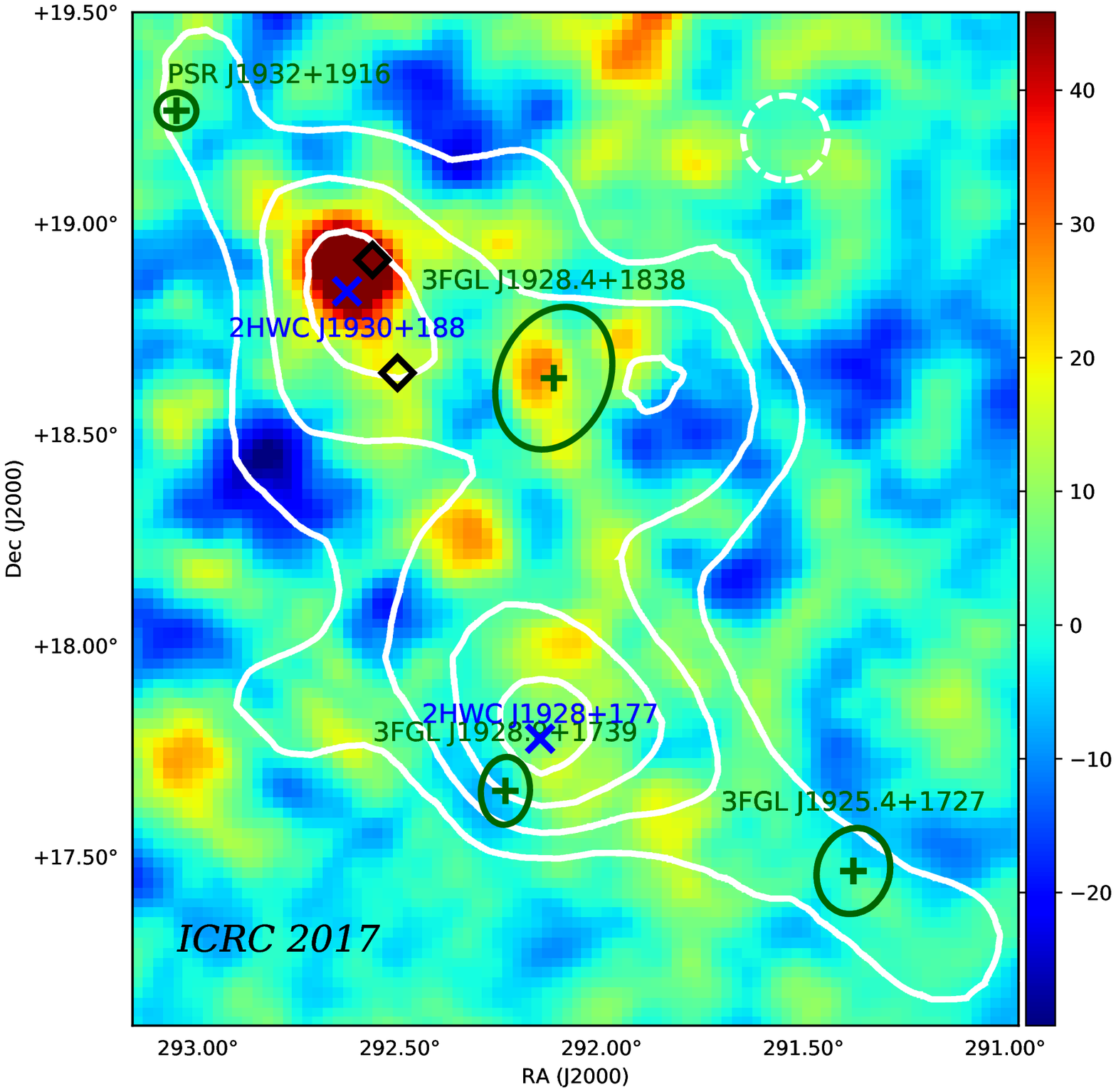}\includegraphics[width=0.47\textwidth]{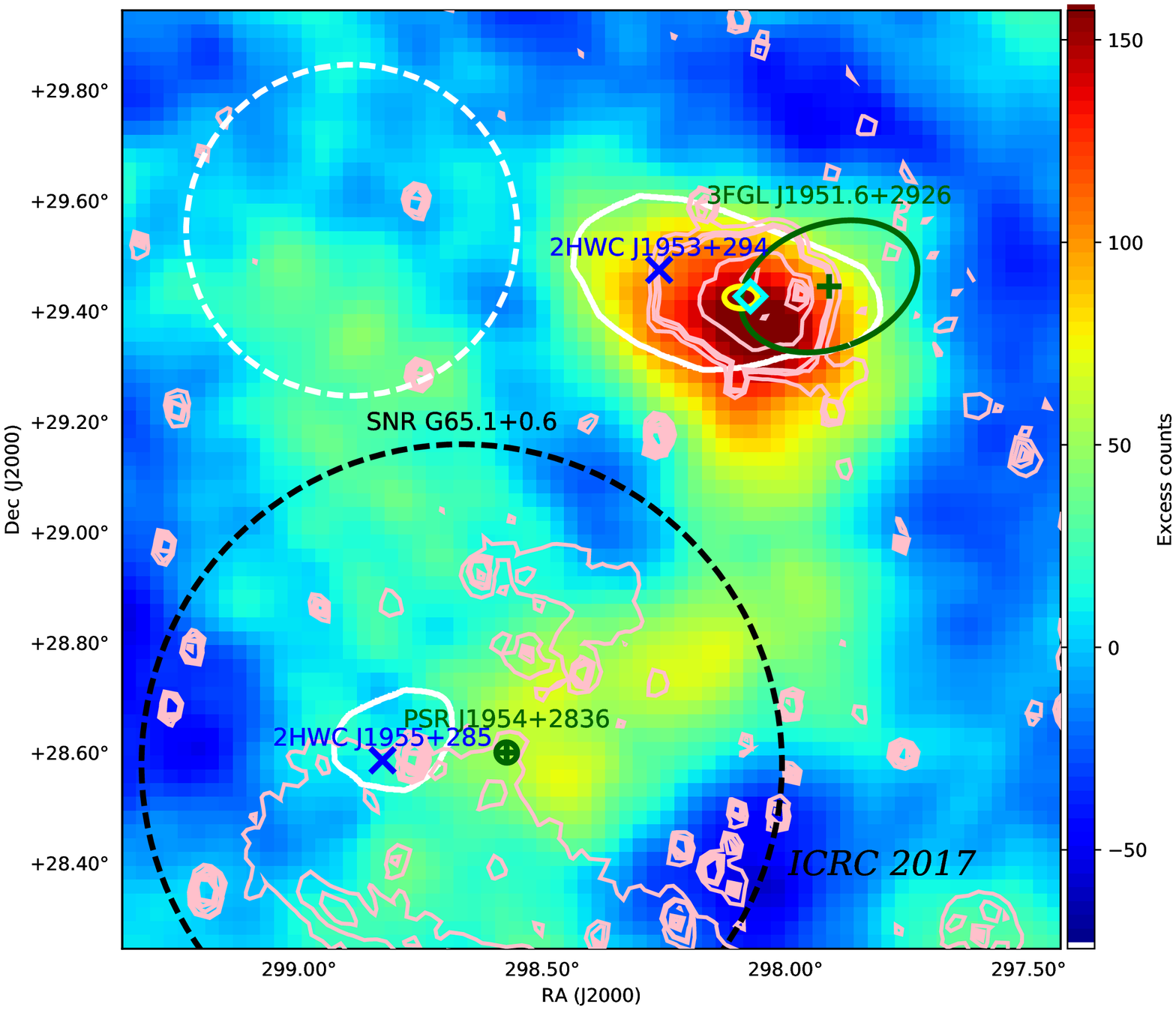} 
  \caption{ (left side) VERITAS gamma-ray counts map of the SNR~G~54.1+0.3 region with point source search cuts.  White contours are HAWC's significance contours of 5, 6, 7, and 8$\sigma$. \textit{Fermi} analysis for this study found a model with two additional point sources best describes the gamma-ray distribution of \textit{Fermi}-LAT data. The locations of the two point sources used in the model are indicated with black diamond markers. For both skymaps, dark green crosses are the locations of 3FGL sources and blue x marks indicate the centroids of two HAWC sources in the region. Also, the $\theta$ cut used for the study is shown as a white dashed circle for both maps.
(right side) VERITAS gamma-ray counts map of the DA~495 region with extended source search cuts. Light pink contours show the radio contours around PWN DA 495 measured by the Canadian Galactic Plane Survey in the 1.42 GHz band~\cite{2003AJ....125.3145T}. White contours are HAWC's significance contours of 5$\sigma$. The cyan diamond is the location of an X-ray compact source. The yellow circle is the centroid of VERITAS. The extension of radio emission from SNR~G~65.1+0.6 is marked with a dashed black line.} \label{Fig:VERITAS_skymap}
\end{figure}

The first region that we discuss in detail contains a TeV source previously identified by VERITAS, VER~J1930+188 \cite{2010ApJ...719L..69A} and a newly identified HAWC source, 2HWC~J1928+177. With 16 hours of additional data taken in the 2015--2016 observing season, VERITAS has a total of 46 hours of exposure in this region. The updated VERITAS counts map overlaid with \textit{Fermi}-LAT and HAWC's information is shown in Figure~\ref{Fig:VERITAS_skymap}. The VERITAS source is associated with the SNR~G~54.1+0.3, a supernova remnant hosting a young, energetic pulsar, PSR~J1930+1852. 

The HAWC source 2HWC J1930+188, coincident with VER~J1930+188, was detected in the HAWC point source search. The centroid of 2HWC J1930+188 is shown in Figure~\ref{Fig:VERITAS_skymap}, and it agrees with the position of the VERITAS source.

The \textit{Fermi}-LAT analysis of this region, searching for the LAT counterpart of HAWC sources, revealed a point source coincident with VER~J1930+188 with a TS value of 26. The non-detection of SNR~G~54.1+0.3 in the 3FGL catalog indicates the possible existence of a low-energy spectral cut-off in the Fermi energy range, but the source is too faint to measure the cut-off in this study.

The spectral index of the HAWC source is softer than that measured by VERITAS. The significance of the difference is 2.4$\sigma$ considering only the statistical errors. 
Extrapolation of the HAWC spectrum to the VERITAS energy range yields an integrated flux that is seven times larger than the VERITAS flux. Although this is still in agreement with VERITAS's measurement within 2$\sigma$ statistical errors, we tested whether HAWC data can be described as a power-law distribution with an exponential cut-off. The result shows that HAWC data can be explained by both a single power-law and a power-law with a cut-off. But, the extrapolation of HAWC's flux for the power-law with a cut-off distribution to VERITAS energies produces an integral flux that is only $\sim$50\% larger than the VERITAS flux, reducing the disagreement between VERITAS and HAWC measurements to 1$\sigma$ statistical error. 


The other HAWC source in the region is 2HWC J1928+177. HAWC reported a similar flux and index for this source as for 2HWC J1930+188. But, unlike 2HWC J1930+188, VERITAS did not detect emission from this source with either the point source search or the extended source search. The angular distance between 2HWC J1930+188 and 2HWC J1928+177 is 1.18$^{\circ}$, which is larger than the PSF of HAWC for energies larger than 1 TeV. Therefore, it is likely that 2HWC J1928+177 is a new gamma-ray source, unassociated with SNR~G~54.1+0.3, and the lack of a VERITAS detection may indicate that the HAWC source has a larger angular extent than the radius of 0.23$^{\circ}$ considered in the VERITAS analysis.

Although there are only two HAWC sources reported in this region, we note that the extension of HAWC's 5$\sigma$ contours covers a larger area than these two sources, as shown in Figure~\ref{Fig:VERITAS_skymap}. It is possible that there are other weak, and possibly extended, TeV gamma-ray emitting sources yet to be clearly identified in this region.

\subsection{DA 495 region}
The second area we discuss in detail is a region around SNR DA 495 (SNR~G~065.7+01.2). HAWC detected two sources with the point source search in this region: 2HWC J1953+294 and 2HWC J1955+285. \textit{Fermi} analysis for the energy range from 10 GeV to 900 GeV did not detect gamma-ray emission in either the point source search or the extended source search. 

After 37 hours of observation, VERITAS reported a confirmation of weak gamma-ray emission nearby 2HWC J1953+294~\cite{2016arXiv160902881H} with an extended source analysis ($\theta<$0.3$^{\circ}$). After this initial report, VERITAS continued observing the source, and accumulated a total of 64 hours of data on the field of view by 2017. VERITAS detected a gamma-ray source in the DA~495 region with 5.2$\sigma$ significance, and we assign the name VER J1952+294.

By assuming a single power-law SED with the measured HAWC index of 2.78, we estimate the integrated flux of VER~J1952+294$^{\ast}$ between 630 GeV and 30 TeV to be (4.46 $\pm$ 2.27$_{\textnormal{stat}}$) $\times 10^{-13} \textnormal{cm}^{-2} \textnormal{s}^{-1}$. This is smaller than the flux measured by HAWC in the same energy range, and the difference between the two measurements is significant at the level of 2.3$\sigma$ when considering only statistical errors. Further study with a deeper exposure will determine whether this is due to a change of spectral index, or due to other reasons, such as the contribution of diffuse emission or a nearby source to the flux measured by HAWC.

The likely counterpart of 2HWC J1953+294 and VER J1952+294$^{\ast}$ is the PWN DA 495. As shown in the Figure~\ref{Fig:VERITAS_skymap}, the emission seen by VERITAS overlaps with the radio contours of DA 495, an X-ray compact source, 3FGL J1951.61+2926 and 2HWC J1953+294. DA 495 has an extended emission in the radio band with emission concentrated in the center. X-ray observations by ROSAT and ASCA revealed a compact central object surrounded by an extended nonthermal X-ray source~\cite{2004ApJ...610L.101A}. The implied blackbody temperature and luminosity, measured by \textit{Chandra}, suggest that the central object is an isolated neutron star. Together with the extended emission surrounding the compact object, this confirms the PWN interpretation of the source~\cite{2004ApJ...610L.101A,2008ApJ...687..505A}.
Based on the low energy break measured in the radio band, \cite{2008ApJ...687..516K} suggested that DA 495 may be an aging PWN with an age of $\sim$20,000 yr.
There is a $\textit{Fermi}$-LAT source, 3FGL J1951.61+2926, coincident with DA~495. However, the extrapolated flux of this source to the HAWC energy range is much smaller than the flux measured by HAWC. It has been suggested that 3FGL J1951.61+2926 is likely associated with the central pulsar of DA~495, although no evidence for pulsations has been identified~\cite{2015MNRAS.453.2241K}.

The second HAWC source in the region, 2HWC J1955+285, is 1$^{\circ}$ away from 2HWC J1953+294. A \textit{Fermi}-LAT detected radio-quiet pulsar, PSR J1954+2836, is located within $2\sigma$ position uncertainty of the HAWC source. But, similar to DA 495, the extrapolation of PSR J1954+2836 to higher energies lies far below HAWC's flux measurement. These sources are within the extent of SNR G 65.1+0.6, a very faint object detected in the radio band. SNR G 65.1+0.6 is a shell-type SNR. The shell strcture can be confined in a circle with a diameter of 70$\arcmin$~\cite{2006A&A...455.1053T}. This is larger than the PSF of HAWC, but the centroid of 2HWC J1955+285 is not coincident with the shell of the SNR. Also, 2HWC J1955+285 is found in a point source search, suggesting that SNR G65.1+0.6 is likely not the counterpart of the HAWC source. Figure~\ref{Fig:VERITAS_skymap} shows that VERITAS sees a region of positive signal around 2HWC J1955+285. The maximum pre-trial significance of this region is 3.2$\sigma$, offset by 0.2$^{\circ}$ from the position of the HAWC source. With the current limited data set, it is unclear whether this is a weak source or simply a statistical fluctuation.

\section{Summary}
Using VERITAS and \textit{Fermi}-LAT, we searched for TeV and GeV gamma-ray counterparts to 13 out of 16 new HAWC sources without clear TeV associations. For eight of those sources, the flux upper limits measured by VERITAS are lower than HAWC's measurements extrapolated to the VERITAS energy range. Among these eight sources, non-detections by VERITAS constrain a point-source hypothesis for five sources with a confidence level of 95$\%$. To resolve the discrepancy between the HAWC and VERITAS measurements, the sources should be angularly extended, or there should be a spectral change in the energy range between VERITAS and HAWC. The extended analysis shows that for two sources among these, 2HWC J1852+013* and 2HWC J1902+048*, the radius of the source should be larger than 0.23$^{\circ}$ to satisfy all of the measurements. These numbers are based on a comparison between the upper limits of VERITAS and the flux estimation of HAWC. However, it is possible that the HAWC flux is overestimated for some of the sources, since the flux estimation has been made with a single point source model for the likelihood analysis without accounting for nearby sources. Unresolved weak diffuse emission over a very large area would also cause an overestimation of the flux.

\addcontentsline{toc}{section}{Acknowledgement}
\small
\section*{Acknowledgements}
\markboth{Acknowledgements}{Acknowledgements}
VERITAS is supported by grants from the U.S. Department of Energy Office of Science, the U.S. National Science Foundation and the Smithsonian Institution, and by NSERC in Canada. We acknowledge the excellent work of the technical support staff at the Fred Lawrence Whipple Observatory and at the collaborating institutions in the construction and operation of the instrument. The VERITAS Collaboration is grateful to Trevor Weekes for his seminal contributions and leadership in the field of VHE gamma-ray astrophysics, which made this study possible.

The \textit{Fermi}-LAT Collaboration acknowledges generous ongoing support from a number of agencies and institutes that have supported both the development and the operation of the LAT as well as scientific data analysis.
These include the National Aeronautics and Space Administration and the Department of Energy in the United States, the Commissariat \`a l'Energie Atomique and the Centre National de la Recherche Scientifique / Institut National de Physique Nucl\'eaire et de Physique des Particules in France, the Agenzia Spaziale Italiana and the Istituto Nazionale di Fisica Nucleare in Italy, the Ministry of Education, Culture, Sports, Science and Technology (MEXT), High Energy Accelerator Research Organization (KEK) and Japan Aerospace Exploration Agency (JAXA) in Japan, and the K.~A.~Wallenberg Foundation, the Swedish Research Council and the Swedish National Space Board in Sweden.

We acknowledge the support from: the US National Science Foundation (NSF); the US Department of Energy Office of High-Energy Physics; the Laboratory Directed Research and Development (LDRD) program of Los Alamos National Laboratory; Consejo Nacional de Ciencia y Tecnolog\'{\i}a (CONACyT), M{\'e}xico (grants 271051, 232656, 260378, 179588, 239762, 254964, 271737, 258865, 243290, 132197), Laboratorio Nacional HAWC de rayos gamma; L'OREAL Fellowship for Women in Science 2014; Red HAWC, M{\'e}xico; DGAPA-UNAM (grants IG100317, IN111315, IN111716-3, IA102715, 109916, IA102917); VIEP-BUAP; PIFI 2012, 2013, PROFOCIE 2014, 2015;the University of Wisconsin Alumni Research Foundation; the Institute of Geophysics, Planetary Physics, and Signatures at Los Alamos National Laboratory; Polish Science Centre grant DEC-2014/13/B/ST9/945; Coordinaci{\'o}n de la Investigaci{\'o}n Cient\'{\i}fica de la Universidad Michoacana. Thanks to Luciano D\'{\i}az and Eduardo Murrieta for technical support.

\end{document}